\documentclass[preprint]{aastex}

\usepackage{graphicx,amssymb,amsmath,multirow,gensymb,lscape}
\usepackage{slashbox,enumitem,arydshln,multirow}
\usepackage{natbib}
\slugcomment{Accepted by AJ}

\shorttitle{Star Formation of Perseus} 
\shortauthors{Mercimek et al.}

\title{An Estimation of the Star Formation Rate in the Perseus Complex}

\author{
Seyma Mercimek$^{1,2}$, Philip C. Myers$^{1}$, Katherine I. Lee$^{1}$, Sarah I. Sadavoy$^{1}$}
\affil{$^{1}$Harvard-Smithsonian Center for Astrophysics, 60
Garden Street, Cambridge, MA 02138, USA; smercimek@ogr.iu.edu.tr}
\affil{$^{2}$Istanbul University, Graduate School of Science and Engineering, Bozdogan Kemeri Cad. 8, Vezneciler-Istanbul-Turkey}

\begin{document}

%%%%%%%%%%%
%%% ABSTRACT%%
%%%%%%%%%%%
\begin{abstract}
We present the results of our investigation of the star-forming potential in the Perseus star forming complex. We build on previous starless core, protostellar core, and young stellar object (YSO) catalogs from {\it Spitzer} (3.6-70 $\mu$m), {\it Herschel} (70-500 $\mu$m), and SCUBA (850 $\mu$m) observations in the literature. We place the cores and YSOs within seven star-forming clumps based on column densities greater than $5 \times 10^{21}$ cm$^{-2}$. We calculate the mean density and free-fall time for 69 starless cores as $\sim$5.55 $\times 10^{-19}$ g cm$^{-3}$ and $\sim$0.1 Myr, respectively, and we estimate the star formation rate for the near future as $\sim$150 {\it M}$_{\odot}$\mbox{Myr$^{-1}$}. According to Bonnor-Ebert stability analysis, we find that majority of starless cores in Perseus are unstable. Broadly, these cores can collapse to form the next generation of stars. We found a relation between starless cores and YSOs, where the numbers of young protostars (Class 0 + Class I) are similar to the numbers of starless cores. This similarity, which shows a one-to-one relation, suggests that these starless cores may form the next generation of stars with approximately the same formation rate as the current generation, as identified by the Class 0 and Class I protostars. It follows that if such a relation between starless core and any YSO stage exists, SFR values of these two populations must be nearly constant. In brief, we propose that this one-to-one relation is an important factor in better understanding the star formation process within a cloud.
\end{abstract}

%%% Keywords

%%% Introduction
\section{Introduction}
%%%
The Perseus molecular cloud is an excellent region to study dense cores and young stellar objects (YSOs) in terms of intermediate and low-mass star formation within isolated and clustered conditions (\citealt{Bal08}, p. 308). Perseus is relatively nearby at 235 pc (\citealt{Hir11}), and many studies have already investigated its dense core and YSO populations using continuum observations (e.g., \citealt{Eno06}; \citealt{Kir06}; \citealt{Eva09}; \citealt{You15}) and molecular line emission (e.g., \citealt{Hat05}; \citealt{Ros08}; \citealt{Fos09}) from different telescopes.  In particular, studies have used these population surveys to classify dense cores as protostellar or starless (e.g., \citealt{Eno08}; \citealt{Jor08}; \citealt{Sad10}).

In addition to studying populations of dense cores and YSOs, continuum observations can also reveal different cloud structures on larger scales. Molecular clouds have a lot of structure, whereas the dense cores and YSOs are often embedded in filaments (e.g., \citealt{And10}) or in moderately dense, larger-scale ($\sim$1 pc) clumps (see Di Francesco et al. 2007, p. 17). Column density (or extinction) maps from continuum observations of entire clouds are excellent tools to trace these structures (e.g., \citealt{Sch15}), and several studies have produced such maps for Perseus (e.g., \citealt{Rid06}; \citealt{Sad14}; \citealt{Zar16}).

Connecting cores to their clumps is important, as cores represent the next generation of stars, and clumps represent the immediate star-forming potential. In this study, we primarily focus on the starless cores and exclude protostellar cores from our main analysis. Thus, we can estimate the next generation star formation rate (SFR) by counting the numbers of starless cores.

Many previous studies have measured SFRs of entire clouds based on their YSO populations, e.g., 57 {\it M}$_{\odot}$ Myr$^{-1}$ in Serpens \citep{Har07}, 6.5 {\it M}$_{\odot}$ Myr$^{-1}$ in Cha II \citep{Alc08}, and 73 {\it M}$_{\odot}$ Myr$^{-1}$ in Ophiuchus \citep{Eva09}. \citet{Eva09} also found an SFR of 96 {\it M}$_{\odot}$ Myr$^{-1}$ for Perseus, which is relatively high compared to similar nearby clouds.

These previous studies generally used different number counts and considered different stages of YSOs.  For example, \citet{Eva09} determined SFRs by taking a mean mass of 0.5 {\it M}$_\odot$ in 2 Myr, which is the typical age of Class II sources. They included all YSO stages for the estimation. We choose a different way to estimate SFR in this study. Working on relations between numbers of YSOs stages and starless cores from clump to clump, we estimate SFR without including the all of the stages. In brief, we do not need to estimate lifetimes and masses of YSOs of all stages.

We describe the core and YSO catalogs in Section 2. We discuss the structures of clumps in Section 3. In Section 4, we present our analysis. Finally, we discuss our results and previous studies in Section 5.

%%%%%
% DATA%
%%%%%
\section{Data}
%%%
We used published catalogs of cores and YSOs at different wavelengths ranging from sub-millimeter (850 $\mu$m) to the infrared (1.25 $\mu$m). In the following section, we explain which data we used for the starless cores, the protostellar cores, and the YSOs.
%%%%%%%%%%%%%%%%%%%%
\subsection{Starless Cores and Protostellar Cores}
%%%%%%%%%%%%
Cores within molecular clouds can be observed in emission at wavelengths > 100 $\mu$m, as these structures are generally cold (< 20 K), small (< 0.1 pc), and dense (> $1 \times 10^{5}$ cm$^{-3}$; \citealt{Di07}, p. 17). Cores that have already contain protostars are identified from the presence of an internal heating source. For example \citet{Dun08}, \citet{Eva09}, and \citet{Kon15} identified protostellar cores via compact 70 $\mu$m emission. Several recent studies identified dense cores and classified them as protostellar or starless using infrared observations of protostars.

First, \citet{Jor08} combined observations from the {\it Spitzer} c2d survey and SCUBA/JCMT submillimeter data. This study classified cores as protostellar if there was an MIPS source located within 15$''$ of the core center. Second, \citet{Eno08} combined {\it Spitzer} c2d data with Bolocam millimeter observations for their sample. They identified protostellar cores using various infrared characteristics such as flux density, infrared colors, and proximities to the core center. Finally, \citet{Sad10} similarly used SCUBA and c2d data to classify cores in five clouds. Their classification scheme identified protostellar cores based on {\it Spitzer} YSOs within a flux-defined boundary, instead of using a fixed circular distance or the core sizes from previous classification studies. For this study, we adopt the core classifications from \citet{Sad10}.

It is obvious that the number of cores between the three methods should be different, as each method focused on different criteria in order to classify cores as protostellar or starless. However, these numbers  vary only slightly between the methods, as is shown in Table \ref{comparemethod}. 

\begin{table}[h!]
\tabletypesize{\scriptsize}
\centering
\caption{Numbers of Starless Cores and Protostellar Cores in Perseus from the literature}\label{comparemethod}
\begin{tabular}{ccc}
\hline\hline
Method & Starless Core & Protostellar Core\\
\hline
Sadavoy &97& 46\\
Enoch&94&49\\
J{\o}rgensen&101&42\\
\hline
\end{tabular}
\end{table}
%%%%%%%%%%%%%
\subsection{Class 0 Sources}
%%%%%%%%%%%%%%%%%
Class 0 sources are the youngest phase of YSOs, when most of the mass is contained within the dense core \citep{And93}. As these sources are deeply embedded, they are detected at wavelengths from mid-IR to submillimeter (etc., \citealt{Eno09}, \citealt{Sad14}). In short, \citet{Eno09} combined Bolocam 1.1 mm and {\it Spitzer} c2d survey data.  \citet{Sad14} identified Class 0 sources more comprehensively from clump to clump, combining {\it Herschel} data at 70-500 $\mu$m and SCUBA data at 850 $\mu$m with the c2d survey. These two studies also selected different criteria to identify Class 0 sources. \citet{Sad14} used the ratio of submillimeter to bolometric luminosity $L_{submm}/L_{bol}$ > 1\% (\citealt{And2000}, p. 59) whereas \citet{Eno09} used $L_{submm}/L_{bol}$ > 0.5\% (\citealt{And93}). Consequently, \citet{Sad14} did not include the extra four sources using $L_{submm}/L_{bol}$ > 1\% in order to avoid taking borderline sources. In addition, they marked whether Class 0 sources were identified as a {\it Spitzer} sources. Thus, we can exclude late stage YSOs (in Section 3.2), which are overlapped with Class 0 sources using this indication (see also their list, \citealt{Sad14}). Although the numbers in two methods are the same (27/28), the sources are different. Accordingly, we used the Class 0 source list for Perseus from \citet {Sad14}.
 
%%%%%%%%%%%%%%%%%%
\subsection{Young Stellar Objects \\ (Class I to Class III)}
%%%%%%%%%%
We used {\it Spitzer} c2d data from \citet{Eva09} to identify the later stages of YSOs. They identified sources as Class I-Flat-II-III using the updated classifications by \citet{Gre94} (see also, \citealt{Lad87}, \citealt{And93}, where sources were identified by their infrared spectral indices, $\alpha$, which measures the SED slope between $\sim2-24$ $\mu$m). The different classes are defined by:

\begin{description}
\item[0/I] $0.3 \leq \alpha$
\item[Flat] $-0.3 \leq \alpha < 0.3$
\item[II] $-1.6 \leq \alpha < -0.3$
\item[III] $\alpha < -1.6$
\end{description}

We adopt the infrared spectral indices given in \citet{Eva09} and correct for extinction in order to identify the quantity of these YSOs in Perseus. In addition, those sources identified as Class 0 from \citet{Sad14} were removed from the Class I list.  
%%%%%%%%
%RESULTS%%
%%%%%%%%
\section{Results}
\subsection{Borders of Clumps and Sources in Clumps}
\begin{figure*}
\centering
\includegraphics[width=13cm]{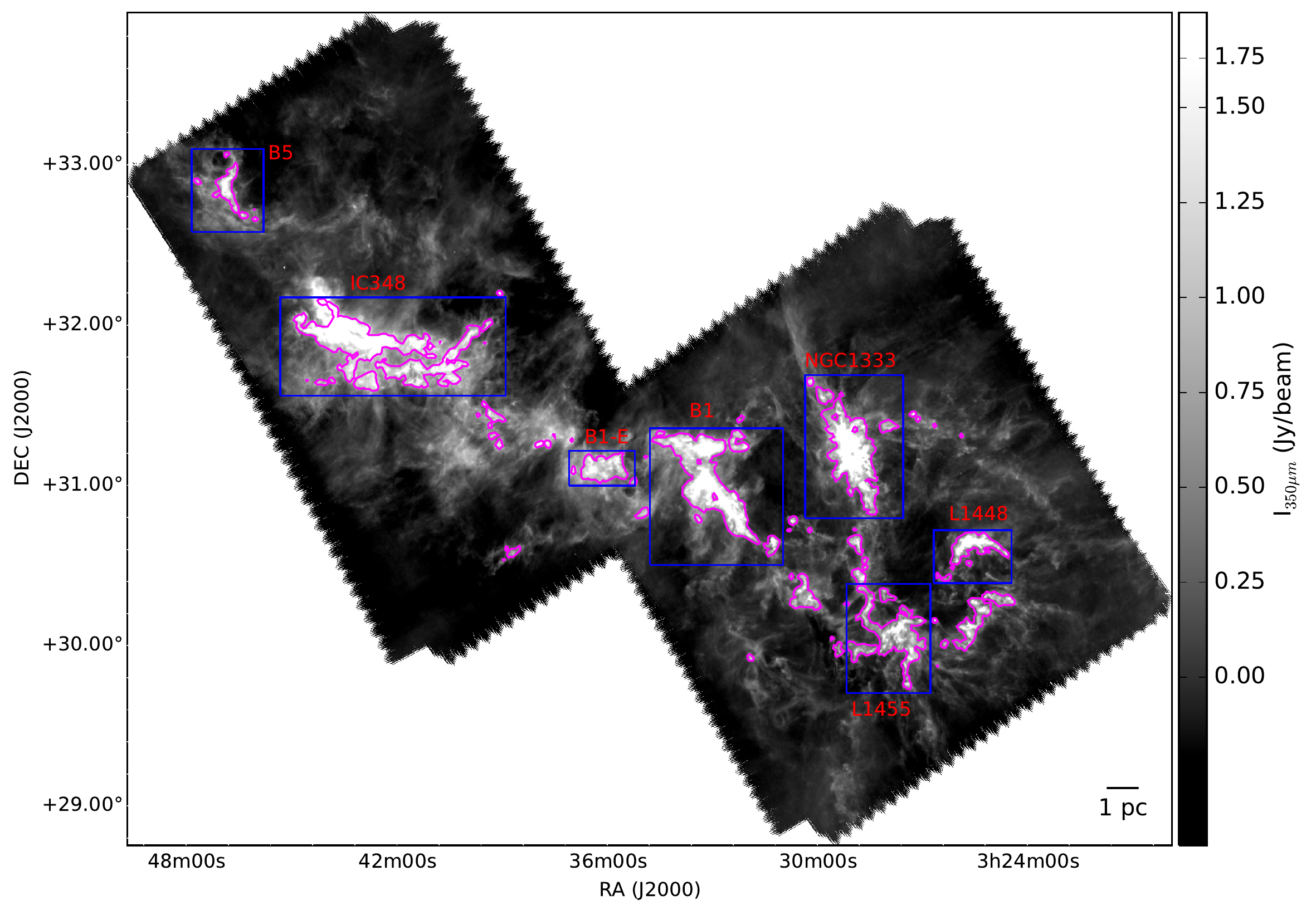}
\caption{{\it Herschel} Level 2.5 image at 350 $\mu$m of Perseus from the Herschel Science Archive. This map corresponds to pipeline-processed data with SPG v9.1.0. The contours are based on $N(H_{2}) \sim 5 \times 10^{21} cm^{-2}$. The blue boxes are distinguish each clump from its neighbors. }\label{border}
\end{figure*}

%%%%%%%%%%%%%%%%
We focus on seven clumps in Perseus, which \citet{Sad14} showed in their Figure 1. They defined these clumps and their boundaries using a fitted {\it Herschel}-derived column density map. The column density threshold of A$_{V}$ $\simeq 7$ mag is proposed as a star formation threshold by \citet{And10}, \citet{Lad10}, and \citet{Eva14}) and is equal to $N(H_{2}) \sim 5 \times 10^{21} cm^{-2}$ (see also, \citealt{Kir06}; \citealt{And10}). The contours of the clumps are based on these thresholds. The names of the clumps are B5, IC 348, B1-E, B1, NGC 1333, L1455, and L1448. We use the same column density threshold of $N(H_{2}) \sim 5 \times 10^{21} cm^{-2}$ $(A_V \simeq 7$ mag) for the clumps in Figure \ref{border}, which shows a 350 $\mu$m map of Perseus from the {\it Herschel} Gould Belt Survey (\citealt{And10}). Each Perseus clumps is illustrated by purple contours, with its name appearing next to it. We exclude sources that are located outside of the clumps in this work.

%Figure 1

We also list protostellar and starless cores in each clump, which is mentioned in Section 2.1. Table \ref{coresnumbers} gives the number of cores.

%%%%%%
\subsection{Clump }
%%%%%%%%%
We considered a core or YSO to be associated with a clump if it is located within the A$_{V}$ = 7 mag contour of that clump from \citet{Sad14} (see also their Figure 1). We define a "source" to be a starless core or a YSO. Table \ref{numbers} lists the quantity of each source in the seven clumps considered in this study. We did not count protostellar cores as sources, because these cores are SCUBA sources that coincided with a {\it Spitzer}-identified as Class 0/I/Flat YSO. Consequently, they would have been double-counted if added to the sources in Table \ref{numbers}. Thus, we only consider the starless cores for this study.

Perseus contains relatively more later-stage YSOs (Class II/III) than young protostars (Class 0/I), as is seen in other studies (e.g., \citealt{Eva09}, \citealt{Dun15}). In addition, most of the sources are found in NGC 1333 and IC 348, which also show high surface densities of sources when compared to the other clumps. Conversely, B1-E and B5 have very few sources despite having >50 {\it M}$_{\odot}$ of material above the column density threshold of star formation. 

\begin{table}[h!]
\centering
\caption{Number of Starless Cores and Protostellar Cores in Each Clump \label{coresnumbers}}
\begin{tabular}{ccc}
\hline\hline
Name of Clump & Starless Core & Protostellar Core\\
\hline
B5 &0& 1\\
B1-E & 0&0 \\
L1448 & 1&3\\
L1455 & 3&4\\
IC 348  & 26&9\\
NGC 1333 &24& 18\\
B1 & 15&8\\
All & 69&43\\
\hline
\end{tabular}
\tablecomments{The values are based on a column density level of $N(H_{2}) > 5 \times 10^{21} cm^{-2}$ and cores from \citet{Sad10}.}
\end{table}

%%%%%%%%
%ANALYSES%
%%%%%%%%
\section{Analyses}

\begin{figure*}
\centering
\includegraphics[scale=0.47]{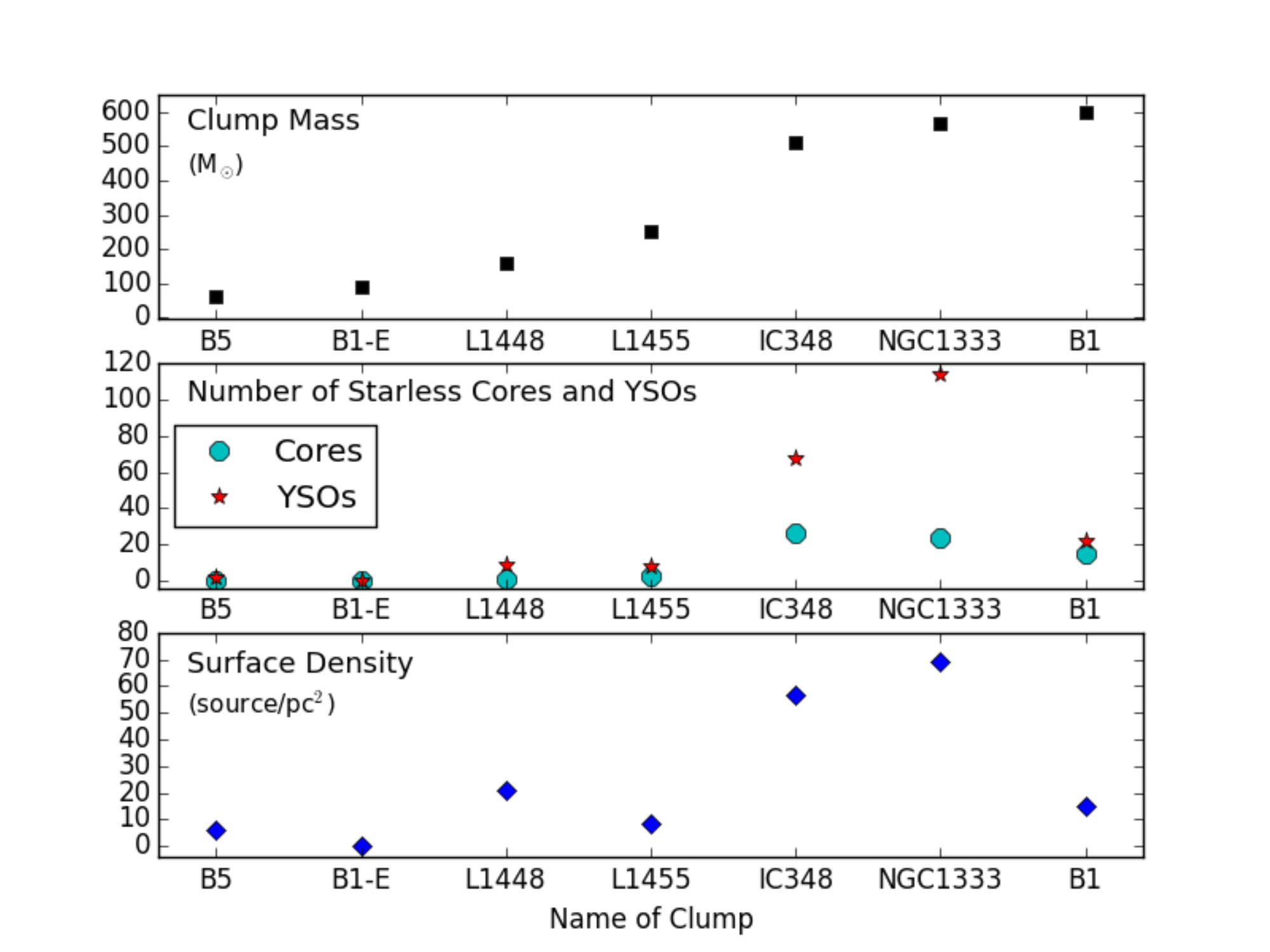}
\caption{Comparison of clump mass (top), source counts (middle), and source surface density (bottom) for each clump in Perseus (see also Table \ref{numbers}).}\label{3plot}
\end{figure*}

Figure \ref{3plot} compares clump masses, the numbers of sources, and surface densities for each of the seven clumps. We order these clumps by mass. In general, we see an increasing trend with respect to those qualities. The clump mass has the added benefit of showing the star formation potential, e.g., the material above $5 \times $10$^{21}$ cm $^{-2}$ (the column density threshold).

According to \citet{Lad10}, the number of YSOs in a cloud should trend with the mass of that cloud above a threshold of A$_{V}$ $\simeq$ 7 mag. Thus, we can make a direct comparison between the Perseus clumps and this expectation. Looking at Table \ref{numbers}, the trend does lean in that direction, but not all clumps appear to follow it perfectly (e.g., IC 348 has more objects than B1, but B1 has more mass above the threshold).

Among all clumps, B1-E has the lowest values in the number of cores and YSOs, while NGC 1333 has the highest values in both. Both IC 348 and NGC 1333 have significant excesses of YSOs including high fractions of Class II and Class III sources compared to starless cores, which suggests they started forming stars early in Perseus compared to the other clumps.  

For the source surface densities, B1 again deviates from this trend, as does L1455, where both seem to have a lower surface densities for their mass. As L1448 and L1455 have a similar number of sources (shown in the middle Figure \ref{3plot}), the lower surface density in L1455 could be attributed to more clustered sources within a larger area. Because our surface densities are averages over the entire clump area for material with A$_{V}$ > 7 mag, highly clustered YSO environments can have underestimated surface densities relative to the larger clumps. 

The deviation in source numbers of B1 suggests that this clump has not yet reached its star-forming potential. Since B1 has such a large star formation potential compared to its source count, the lack of objects suggests it is still very early in its star formation process. This region has very few "evolved YSOs" (and no Class III sources, see Table \ref{numbers}), which is indicative of a young population compared to NGC 1333 and IC 348 (see also \citealt{Bal08}, p. 308).

As indicated in Table \ref{numbers}, B1-E does not show star formation activity. As B1-E is likely very young, it may still need time to form populations of dense cores and protostars (e.g., \citealt{Sad15}).

These same increase trends between those properties of each clump show that both surface densities and number of sources with clump masses increase following the order. For example, there is a linear correlation with coefficient value of 0.7 between clump masses and surface densities from clump to clump. On the other hand, this value could be higher if the surface density of B1 was not  as low, as is shown in Figure \ref{3plot}. In the cases of either the absence or high surface density of B1, the correlation is approximate 1, which is a statistically robust linear relation.

%Figure 2

%Figure 3
\begin{figure*}
\centering
\includegraphics[scale=0.47]{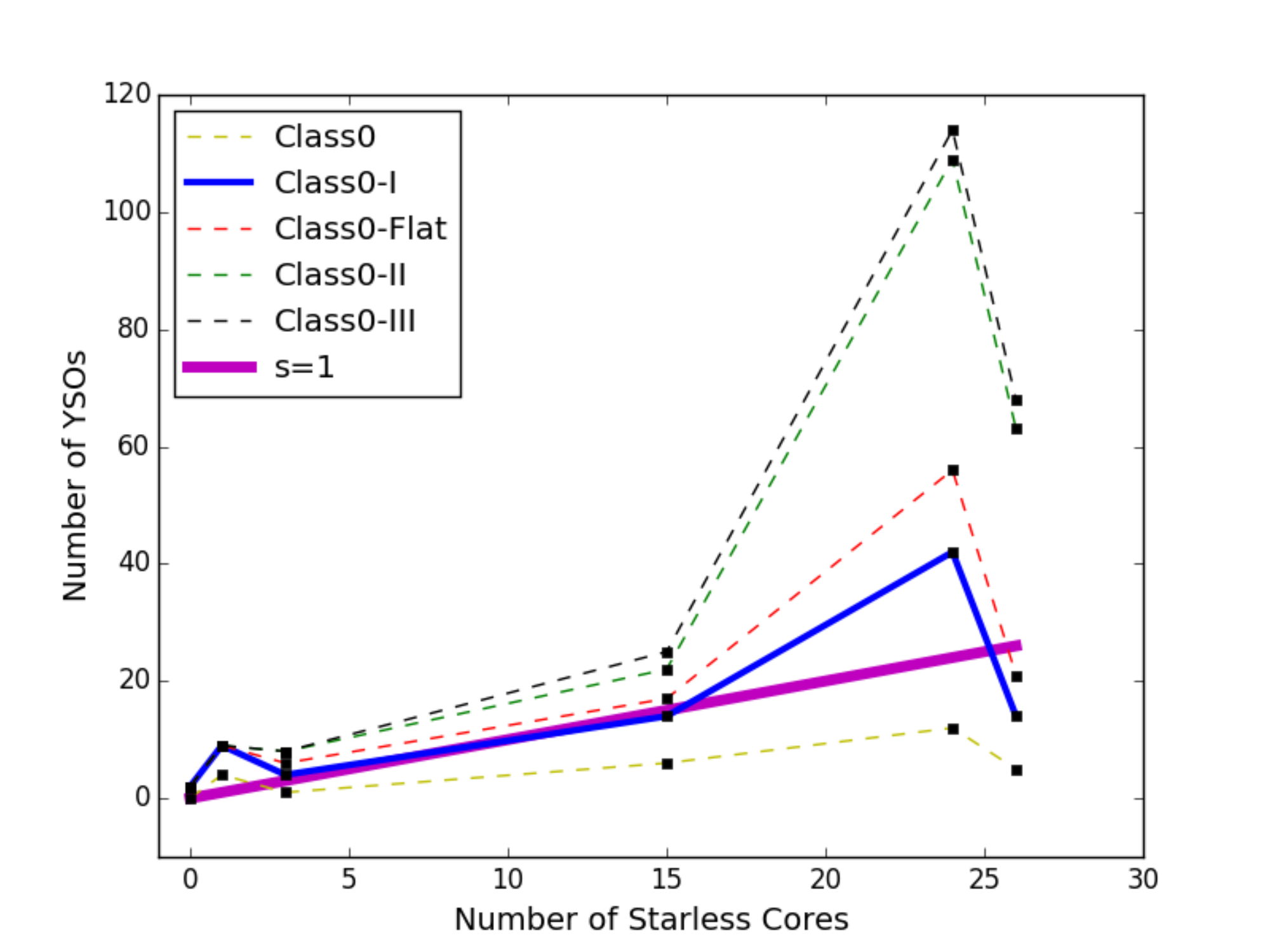}
\caption{Number of starless cores vs. the number of YSOs for the Perseus clumps. From the bottom line to the top line, the order of the YSO classes is: Class 0, Class 0+I, Class 0+I+Flat, Class 0+I+Flat+II, and finally Class 0+I+Flat+II+III. The purple line illustrates a slope value of 1. This represents a one-to-one relation where the number of starless cores equals the number of YSOs. The blue line, Class 0+I, has the closest slope value to purple line. Clumps are shown by black squares and are ordered as B5, L1448, L1455, B1, NGC 1333, and IC 348, in order increasing number of starless core (not containing B1-E).}\label{fits}
\end{figure*}

\begin{deluxetable}{llcccccccc}
\tabletypesize{\scriptsize}
\tablewidth{0pt}
\tablecolumns{10}
\tablecaption{Number of Sources in Each Clump\label{numbers}}
\tablehead{
\colhead{Name of Clump} &
\colhead{Starless\tablenotemark{a}} &
\colhead{Class 0\tablenotemark{b}} &
\colhead{Class I\tablenotemark{c}}&
\colhead{Class Flat\tablenotemark{c}}  &
\colhead{Class II\tablenotemark{c}}  &
\colhead{Class III\tablenotemark{c}}  & 
\colhead{Area\tablenotemark{b}} &
\colhead{Surface Density } &
\colhead{Clump Mass \tablenotemark{b}} \\
	&
Core	&
 	&
	&
	&
	&
	 &
(pc$^{2}$) &
(source pc$^{-2}$)&
({\it M}$_{\odot}$) 
}
\startdata
\hline
B5 &0& 0 & 2 & 0 & 0 & 0 & 0.32 & 6.3 &  62\\
B1-E & 0&0 & 0 & 0 & 0 & 0 & 0.57 & 0 &  88\\
L1448 & 1&4 & 5 & 0 & 0 & 0 & 0.48 & 20.8 & 159\\
L1455 & 3&1 & 3 & 2 & 2 & 0  & 1.3 & 8.5  & 251\\
IC 348  & 26&5 & 9 & 7 & 42 & 4 & 2.9 & 32 & 511\\
NGC 1333 &24& 12 & 30 & 14 & 53 & 5  & 2.0 & 69.0 & 568\\
B1 & 15&6 & 8 & 3 & 5 & 0 & 2.5 & 14.8 & 598\\
All & 69&28 & 57 & 26 & 102 & 9  & 10.7 & 151.4 & 2237\\
\enddata
\tablecomments{The values are based on column density level of $N$(H$_2$) $ > 5 \times 10^{21}$ cm $^{-2}$.}
\tablenotetext{a}{From \citet{Sad10}.}
\tablenotetext{b}{From \citet{Sad14}.}
\tablenotetext{c}{YSO classes are determined from infrared spectral indices from \citet{Eva09}.}
\end{deluxetable}

%Figure 4

\begin{figure*}
\centering
%\begin{turn}{90}
\includegraphics[width=14cm]{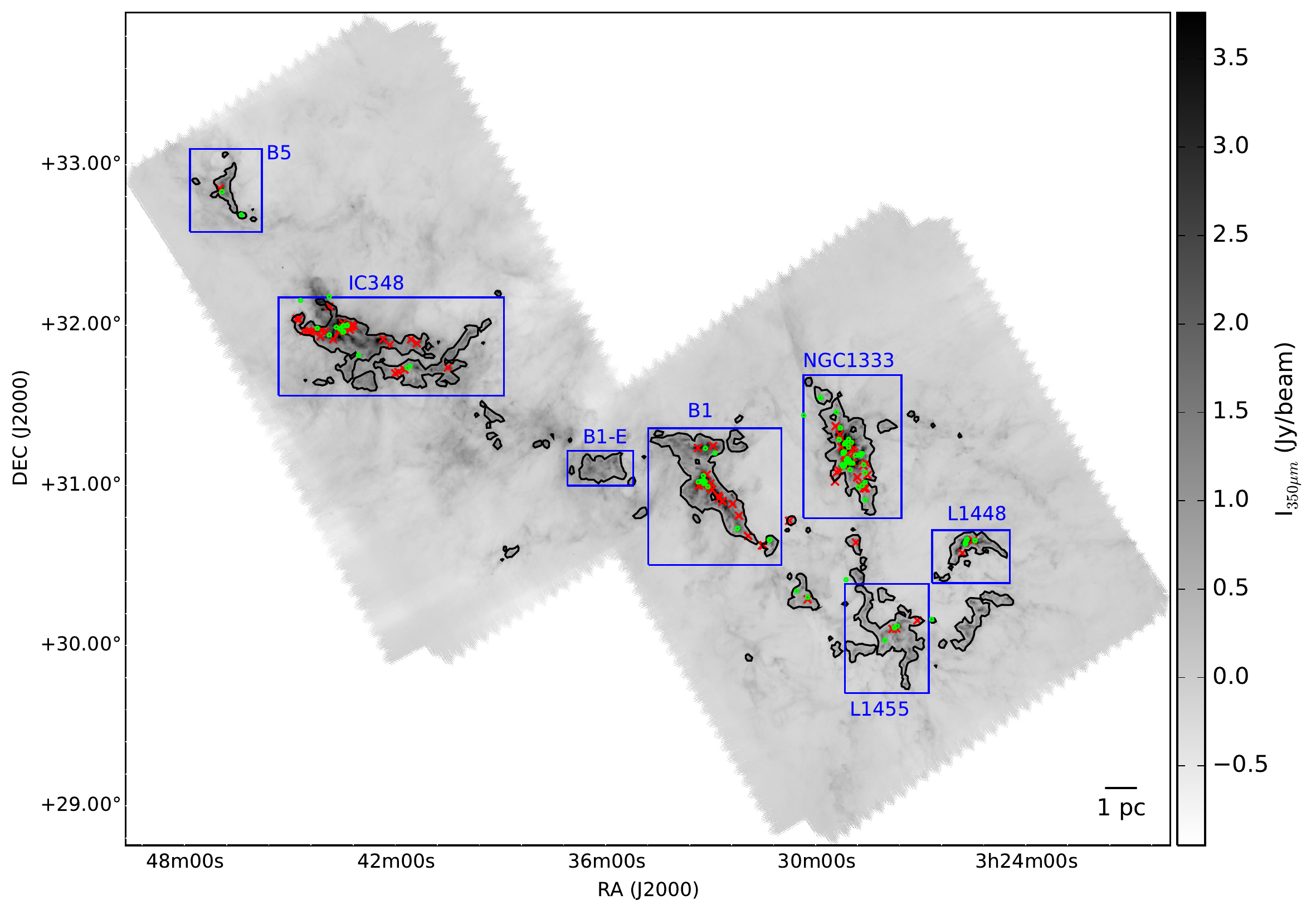}
%\end{turn}
\caption{The seven Perseus clumps with their starless cores and Class 0+I protostars depicted here as red crosses and green circles, respectively. The background images are the same in Figure 1.}\label{zoomingPerseus}
\end{figure*}

%%%%%
\subsection{Relationship Between Starless Cores and YSOs}
%%%%%

Starless cores are the precursors to stars and represent the next generation of stars in molecular clouds. After they collapse, starless cores evolve into Class 0, Class I, Class Flat, Class II, and Class III, respectively. If starless cores reflect the star formation activity of the future, then the YSOs can be considered as the star formation activity of the recent past. By comparing these populations, we can determine how the star formation activity in Perseus is evolving.

Figure \ref{fits} compares the number of starless cores with the numbers of YSOs in each clump. We show distributions for cumulative YSO classifications starting from the youngest stage, Class 0. The thick, solid blue line shows the combined Class 0 and Class I source counts and has the best-fit slope of $\sim$1.04 (linear least square). This is similar to the purple line that illustrates a one-to-one relation between the starless cores and YSOs. Additionally, the correlation coefficient r between Class 0/I and starless cores is approximately 0.8, which indicates a strong linear relation. In other words, the starless cores appear to follow the number of young protostars (Class 0 + Class I) for all clumps in Perseus. In brief, we show with Figure \ref{fits} shows which YSO stage(s) are numerically similar to starless cores using values of their fit slopes.

Figure \ref{zoomingPerseus} shows starless cores and Class 0/I protostars in the each clump. The two populations show similar numbers and appear to follow the densest material in the cloud. More evolved YSOs may have moved from their original places of birth, as was recently suggested in Orion (\citealt{Meg16}, \citealt{SG16}). Because starless cores are considered the progenitors of stars, this similarity suggests that the SFR should be relatively similar between the current generation of young protostars and the next generation. 

In order to examine that most of the starless cores in the Perseus clumps collapse, the Bonnor-Ebert mass of each core is estimated. If the Bonnor-Ebert sphere mass is less than the core mass, the starless core can collapse and form a new star (\citealt{Spi68}, p. 44). The critical mass values are defined by \citet{Ebe55} and \citet{Bon56}:

\begin{equation}
M_{crit} = 1.18 \frac{c_{s}^{4}}{G^{3/2}P_{BE}^{1/2}}, \label{BE}
\end{equation}

\noindent where {\it c} is the isothermal speed of sound. P$_{BE}$ = $\rho_{o}$c$_{s}^{2}$ is the boundary pressure of a core, where $\rho_{o}$ is surface density of a core, assuming the core temperature is 10K. On the other hand, we can estimate the mean density of starless cores ($\rho_{mean}$) with Equation \ref{density}. We calculated the Bonnor-Ebert sphere masses  of the starless cores in Perseus (for the Bonnor-Ebert sphere, the ratio of mean density to surface density is 2.43). 

\begin{figure*}
\centering
\includegraphics[scale=0.38]{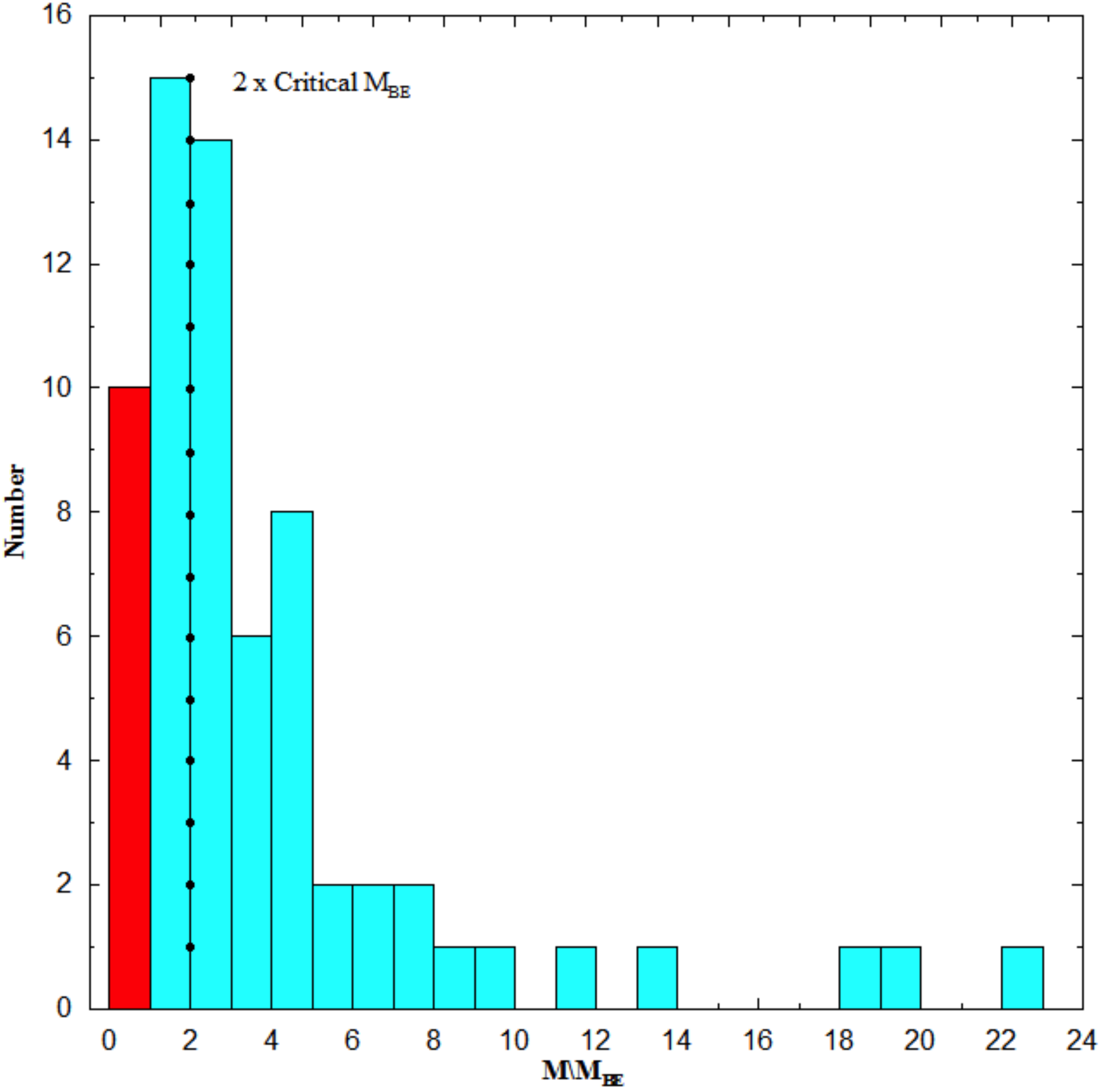}
\caption{Histogram of the ratio of the starless core mass to the Bonnor-Ebert mass. Cyan columns depict ratios greater than 1, and the red column shows the ratios less than 1. The dotted line highlights 2 x M$_{BE}$.}\label{BEhistogram}
\end{figure*}

Figure \ref{BEhistogram} shows the ratio of core masses to Bonnor-Ebert sphere masses. We know that if this ratio is greater than 1, cores are unstable and start collapsing, as mentioned above. Only 10 of the 69 starless cores (red column) have ratio less than 1. In the Bonnor-Ebert Model, all of the others (59 starless cores) are quite massive for stable equilibrium. We've also excluded three of the most massive cores from the histogram, as their ratios exceeded 20. Even if we consider typical maximum mass with an uncertainty factor of 2 (the dotted line in Figure \ref{BEhistogram}), we see substantial cores with masses greater than 2 times the Bonnor-Ebert masses. \citet{Dun16} worked on the stability of cores in Chameleon. Their study focused on whether starless cores in Chameleon are stable or not. According to their result, cores between 0.5 and 2 times the Bonnor-Ebert mass (0.5 M$_{BE}$ < M < 2 M$_{BE}$)  are potentially unstable. While most of the cores in Chameleon are stable (core masses are under 0.5 M$_{BE}$), the majority of starless cores in Perseus are primarily unstable (core masses are above 2 M$_{BE}$). In the light of such information, we can assume that starless cores in Perseus collapse to make next generation of stars.

Assuming that the numbers of starless cores and protostars are equal, we estimate the future SFR.

\begin{equation}
\dot{M}_{stars,last} = \dot{M}_{stars,next}
\end{equation}

\noindent where $\dot{M}$$_{stars}$ is rate of stellar mass gain. Assuming steady state, we can eliminate times, resulting in,
\begin{equation}
{M_{stars,last}} = {M_{stars,next}}, \label{equalM}
\end{equation}
\noindent where M$_{stars,last}$ and M$_{stars,next}$ represent the masses of the protostars and the starless cores, respectively. Hereafter, we use M$_{next}$  for M$_{stars,next}$ and M$_{last}$ for M$_{stars,last}$. Using Equation \ref{equalM}, the total mass of young protostars can be calculated,  because we already have a value for the total mass of starless cores (from \citealt{Sad10}), which is $\sim$190 {\it M}$_{\odot}$ for 69 starless cores. Not all of the core mass goes into the next generation of stars. For instance, there is some efficiency factor (\citealt{Alv07}) thar must be considered. For the protostellar mass, \citet{WK06} estimated the average stellar mass for standard initial mass functions (IMFs) between 0.01 and 150 {\it M}$_{\odot}$ as,
\begin{equation}
\overline{m}_{IMF}= 0.36\:\mbox{M$_{\odot}$}. \label{av}
\end{equation}

We assume that the average final masses of Class 0 and Class I YSOs are equal to $\overline{m}$$_{IMF}$, which means

\begin{equation}
M_{last} = N_{last}\:\overline{m}_{IMF}. \label{equalM}
\end{equation}

For 85 protostars, the expected mass of protostars is

\begin{equation}
\mbox{M$_{last}$} \cong 31\: \mbox{M$_{\odot}$}. \label{Mass}
\end{equation}

With these numbers and masses of cores and stars, "core-star number efficiency" and "core-star mass efficiency" can be estimated as 

\begin{equation}
\epsilon_{csn} = \frac{N_{last}}{N_{next}} \sim 1.2 \: \: \mbox{and}
\end{equation}

\begin{equation}
\epsilon_{csm} = \frac{M_{last}}{M_{next}} \sim 0.16. 
\end{equation}

Our analysis primarily assumes that all of our protostars are single systems. Nevertheless, many of the Perseus protostars are binaries or higher order multiples (see \citealt{Lee15}, \citealt{Tob16}). If we were to consider multiple systems, then both $\epsilon_{csm}$ and $\epsilon_{csn}$ would increase. Therefore, our values should be taken as lower limits (within the assumptions of this study) to achieve steady state. Thus, we find that at least 1.2 stars will be produced from a core and that at least 16\% of the core mass will be turned into stars.

%%%%%%%%%%%
%FREE FALL TIME%
\subsection{Calculation of Free-fall Time and Weighted Mean Lifetime}

If a cloud mass exceeds its virial mass, it should become unstable and collapse unless there is another mechanism of support, such as turbulence or magnetic fields (\citealt{Bod11}, p. 68). Without additional support, the cores collapse under self-gravity in a free-fall time,
\begin{equation}
t_{ff} = (\frac{3\pi}{32G\rho})^{1/2} \label{tff} ,
\end{equation}
\noindent where $\rho$ is the mean core density. We estimate these densities from the effective radii and masses of the starless cores (from \citealt{Sad10}), assuming spherical symmetry. We find a mean starless core density

\begin{equation}
\overline{\rho} = 5.55 \times 10^{-19} \:\mbox{g cm$^{-3}$}. \label{density}
\end{equation}

\begin{figure*}
\centering
\includegraphics[scale=0.47]{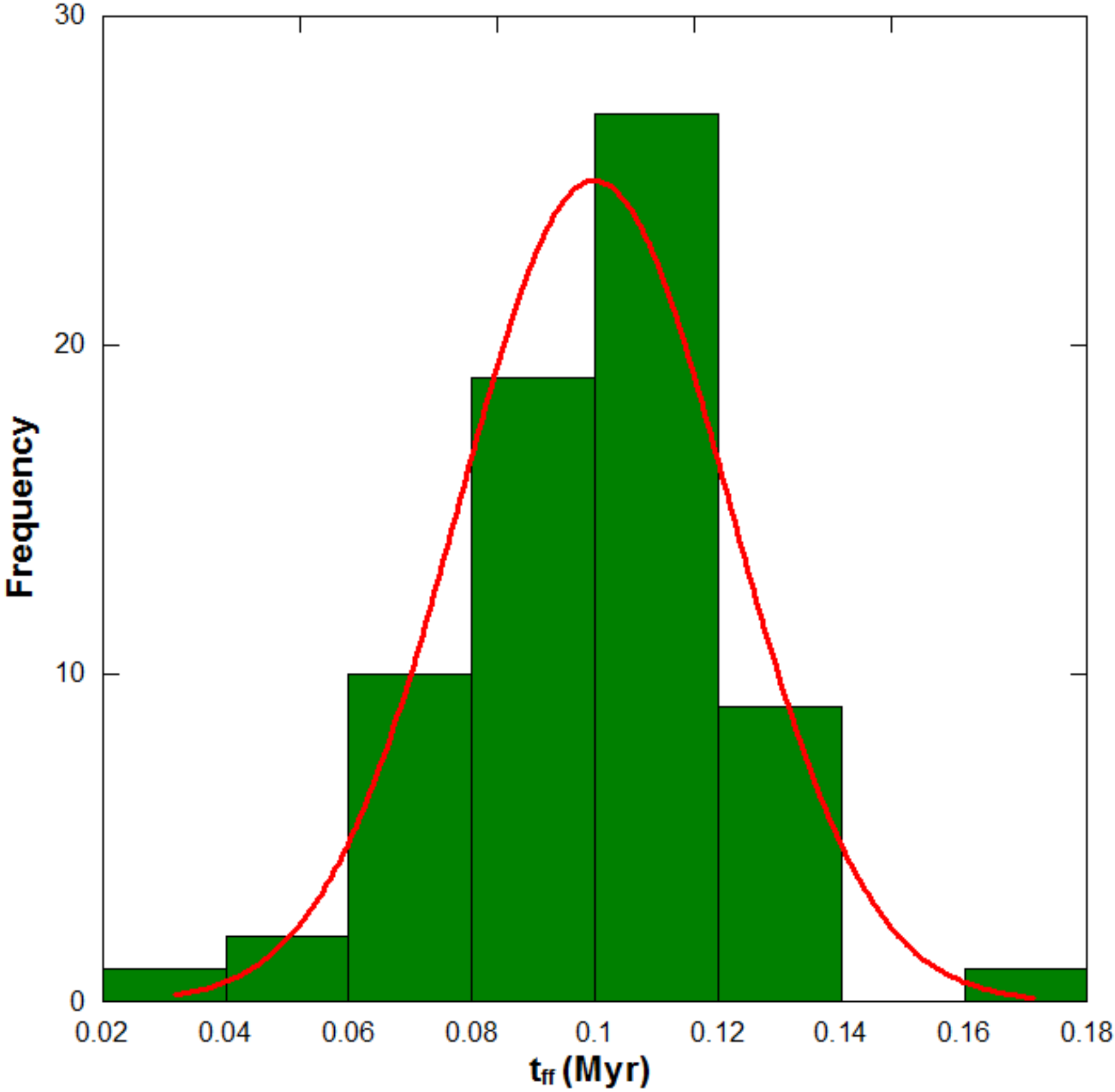}
\caption{Histogram of the ratio of starless core free-fall times. The best-fit Gaussian has $\mu, \sigma$ values of 0.10 Myr and 0.02 Myr, respectively.}\label{histogram}
\end{figure*}

Figure \ref{histogram} shows the distribution of free-fall times for all 69 starless cores, using Equation (\ref{tff}), and their average densities. We find a narrow distribution of free-fall times with an average of

\begin{equation}
\overline{t}_{ff} = 0.10\:\mbox{Myr.}
\end{equation}

%Figure 5

Thus, we expect the starless cores to collapse in $\sim$0.10 Myr, if they are all nearly virialized.

The starless cores will only collapse on a free-fall time if they are unstable. For example, high internal turbulence could delay their collapse by several free-fall times (e.g., \citealt{Nak98}). Nevertheless, if the starless core and protostars are in a steady state, we can assume that the starless core lifetime equals the protostellar core lifetime. We can estimate the protostellar core lifetime by a weighted mean time ($\overline{\tau}$),

\begin{equation}
 \overline{\tau} = \omega_{0}\tau_{0} + \omega_{I}\tau_{I},
\end{equation}

\noindent where $\omega$$_{0}$ and $\omega$$_{I}$ are correspond to the fraction of Class 0 and Class I protostars, respectively (e.g., N$_{Class0}$ / N$_{Class0+ClassI}$ and  N$_{ClassI}$ / N$_{Class0+ClassI}$) and $\tau_{0}$ and $\tau_{1}$ are the lifetimes of the Class 0 and Class I stages. We adopt the lifetimes from \citet{Eva09}, which were 0.10 Myr and 0.44 Myr for Class 0 and Class I YSOs, respectively. Thus, we find a weighted mean lifetime of the Perseus young protostars of
\begin{equation}
\overline{\tau} = 0.33 \:\mbox{Myr}. \label{time}
\end{equation}

Accordingly, we can assume the starless cores will form the next generation of protostars in 0.33 Myr. In units of the free-fall time, we find, 

\begin{equation}
\frac{\overline{\tau}_{next}}{\overline{\tau}_{ff}} = 3.3.
\end{equation}

Thus, actual formation time is less efficient than pure free-fall time by a factor of  $\epsilon_{cst}$ = 3.3, assuming steady state.

As a result, the average starless core makes approximately 1.2 stars in 3.3 free-fall times.

\subsection{The SFR in Perseus}

The SFR corresponds to the star-forming mass as a function of lifetime. Similar to the weighted mean lifetime (Equation \ref{time}), we found the weighted mean inverse lifetime as,

\begin{equation}
 \overline{\tau^{-1} } = \omega_{0}\tau_{0}^{-1} + \omega_{I}\tau_{I}^{-1},
\end{equation}

\begin{equation}
\overline{\tau^{-1} } = (0.21 \:\mbox{Myr})^{-1} . \label{timeSFR}
\end{equation}

According to this, we estimate the SFR, using our number counts of protostars and their weighted lifetime. Following from Equation \ref{Mass} and Equation \ref{timeSFR}, the weighted SFR is
\begin{equation}
\overline{SFR} = M_{stars}\:\overline{\tau^{-1}},
\end{equation}

\begin{equation}
\overline{SFR} \cong 150\: \mbox{M$_{\odot}$}\:\mbox{Myr$^{-1}$}.
\end{equation}

%Figure 6

%%%%%%%%%
%DISCUSSION%
%%%%%%%%%
\section{Discussion}

We estimated the SFR in Perseus using observations of young protostars and starless cores. We found that the SFR in Perseus is 150 {\it M}$_{\odot}$ Myr$^{-1}$ using M$_{stars}$ and $\overline{\tau}$. If our SFR included close binary pairs, it could be higher. For example, if we instead take the starless cores, assuming 1.2 stars form per core at an efficiency of 16\%, we get 174 {\it M}$_{\odot}$ Myr$^{-1}$. In addition, 1.2 protostars from each core implies that there would be a binary fraction of 20\% (one out of five cores will have a companion). The binary fraction from \citet{Tob16} is 40\%, so the SFR could be a bit higher. Of course, the SFR values of clouds in the nearby Gould Belt are not as high as those in giant molecular clouds toward the Galactic plane, where high-mass stars are forming (e.g., \citealt{Ven13}).

The SCUBA data used here will primarily pull out dense objects that are more likely to form stars in the near future. Following that, there is also the problem of starless core masses (single temperature, single dust properties, etc.); however, in the absence of robust temperature or more data, we cannot constrain these properties. With our assumptions (fixed dust temperatures and dust opacities for all starless cores, fixed stellar masses for the protostars, and a constant SFR between the protostars and the starless cores), our result $\epsilon_{csm}$ = 0.16 is similar to the value $\epsilon_{csm}$ = 0.17 found in Perseus by \citet{Jor08} using a similar method. These values are lower than the estimates $\epsilon_{csm}$ = 0.3 in the Pipe Nebula by \citet{Alv07} and the $\epsilon_{csm}$ = 0.4 in Aquila by \citet{Kon15}, by matching IMF with core mass functions. If we included multiple stellar systems or decreased the core mass (either by a higher dust temperature or a higher dust opacity), $\epsilon_{csm}$ would increase. Therefore, the differences between our $\epsilon_{csm}$ value and values of these studies could be due to these assumptions.

We found a near one-to-one relation between starless cores and protostars. This suggests that the SFR in Perseus may be constant (until the next generation) if all of the starless cores in this sample are likely to form stars in the near future. Because SCUBA observations are good at picking out the densest cores, whereas {\it Herschel} observations will also select more diffuse objects, it is reasonable to assume the SCUBA cores are likely to form new stars. According to Bonnor-Ebert criterion (\citealt{Spi68}, p. 44), core condition can be understood regardless of whether the core collapses to form new stars or not. If the ratio of the core mass to the critically Bonnor-Ebert sphere mass exceeds 1, then there is no equilibrium available, and the sphere is unstable against collapse. Subsequently, the core will collapse under its gravity.  Assuming the core temperature is 10 K, nearly all (59 out of 69) starless cores are unstable in this work. Rates which go up to 142 are quite high. This means that core criteria (\citealt{Sad10}) with SCUBA observations tend to select dense cores. 

%Figure 7
\begin{figure*}[h!]
\centering
\includegraphics[scale=0.45]{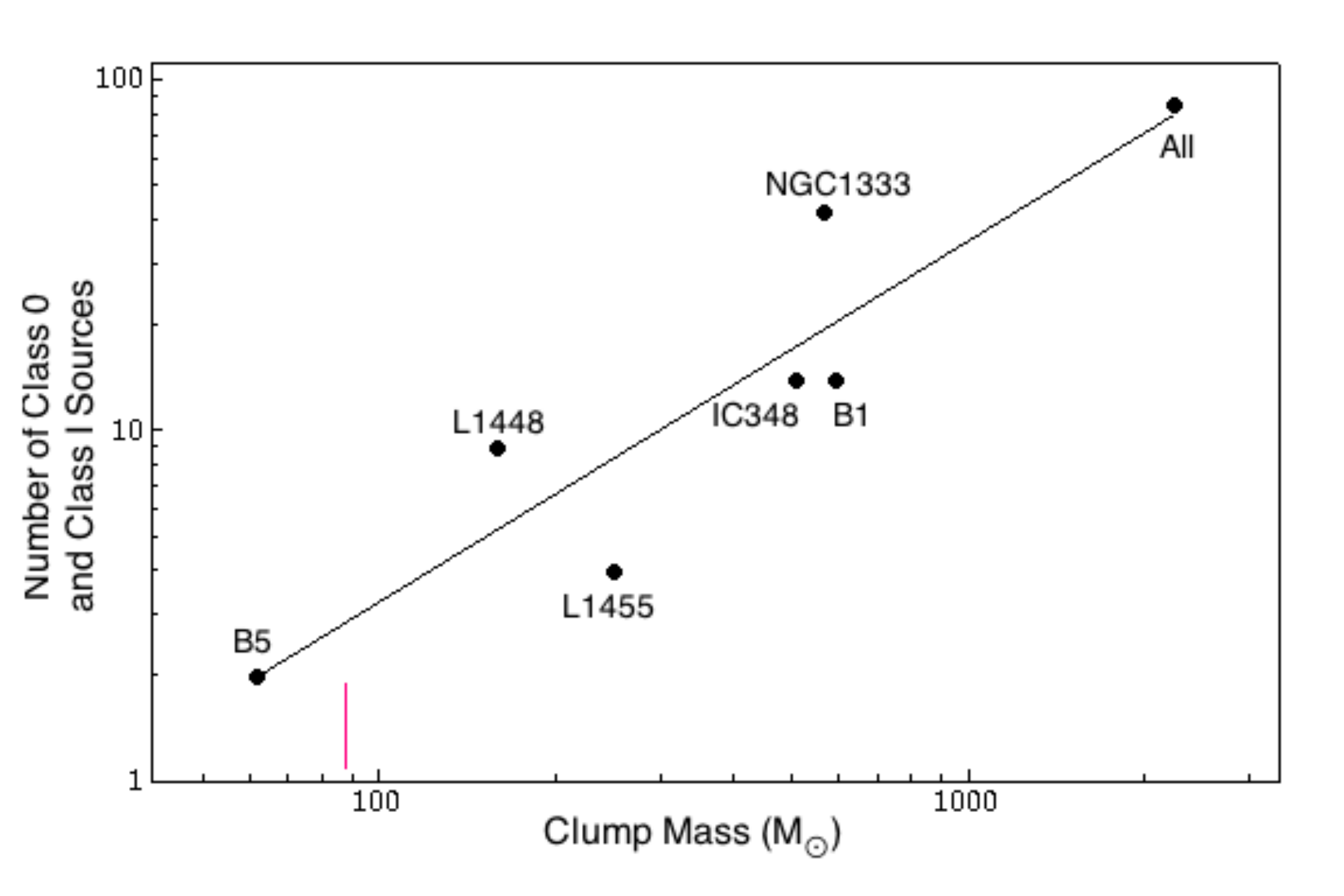}
\caption{Number of protostars vs. clump masses. We show titled clumps with the included power-law tail (slope value of 1.08). The solid pink line shows where B1-E should be located.}\label{ProtoClump}
\end{figure*}

We acknowledge that the main assumption of this study which states that a one-to-one relation can be used if most of starless cores in a cloud are unstable. Nevertheless, the majority of studies demonstrate that starless cores in a cloud are not unstable to collapse, using various methods to do so. For example, \citet{BelA}, \citet{Tsi15}, and \citet{Dun16} found that most of the starless cores in Cha I are stable, and cores cannot form stars. Following that, nearly all of the starless cores are stable in Orion B  (\citealt{Kir16}). On the other hand, one recent studyfound that more than half of starless cores in the Aquila molecular cloud complex are unstable (\citealt{Kon15}). Considering all of these results, Perseus appears to be a special molecular cloud in terms of starless cores. Most of the starless cores in the cloud are unstable, and they will turn into stars in the near future.

We also compared the number of protostars with their clump properties. Figure \ref{ProtoClump} shows that the protostellar counts for the individual clumps (except B1-E because it has no sources) and the full Perseus cloud follow a power-law relation (slope value of 1.08) with clump mass similar to what \citet{Lad10} found for entire clouds. We find more scatter than the \citet{Lad10} relation, which considered YSOs of all stages. A similar plot of the clumps in Perseus is highly dependent on the population age. For example, both B1 and B1-E have relatively few sources of all stages given their clump masses (see Figure \ref{3plot}), and both are considered to be young regions. For individual clumps, it is possible that the population age plays a role, e.g., both B1 and B1-E have not had the chance to fully reach their star formation potential. Thus, in Figure \ref{ProtoClump}, we consider just the young population (protostars), and the relation is quite good, even for B1 (excluding B1-E, which is still too young). The solid pink line in Figure \ref{ProtoClump} shows where B1-E should be located on this relation. We expect this clump to have about two protostars at the current epoch, whereas it has none.

This study shows that protostars and starless cores have nearly equal numbers in all seven clumps. Assuming the starless cores represent the next generation of stars and the Class 0/I sources represent the most recent generation of stars, the similarity between these two populations suggests a nearly constant SFR across each of the individual clumps in Perseus. In addition, we find a power-law relation between the number of protostars in individual clumps and the dense gas mass of those clumps (except B1-E). We also find that this relation scales with the total cloud protostellar count and dense gas mass. Nevertheless, we do not see this same relation with clump mass when we also consider the more evolved YSOs. Using this approach, relations between sources and star formation can be used to study for other clouds.

\acknowledgments
This study was supported by research funding through AAG84 group (\"{U}sk\"{u}dar American Academy), which consists of benefactors who support successful female students in Turkey. We would like to thank to some group members who include Ay\c{s}{\i}n Akbarut, Zeynep Ata\c{c}, Filiz Diniz, Nil Canal, Zeynep Ece, Sara Angat, Asl{\i}han Alt{\i}nkaya, Neslihan Alt{\i}nkaya, G\"{u}nay \c{S}en, Funda C{\i}b{\i}ro\u{g}lu, Aysel Kartal, and M\"{u}ge Tuna. We also thank to Dr. Michael M. Dunham (Harvard-Smithsonian Center for Astrophysics, Cambridge, USA) for useful suggestion and Prof. Y\"{u}ksel Karata\c{s} (Istanbul University, Istanbul, Turkey) for his support during this work. 

\appendix
\section{Tables of Sources For Each Clump}

We list cores and YSOs for each clump in the Perseus in Table \ref{Starless} to Table \ref{ClassIII}. They are selected according to the contour of column density level as $5\times 10^{21} cm^{-2}$. The ID field shows the number of sources in the clumps. The properties of starless cores and protostellar cores are obtained from \citet{Sad10} and one protostellar core in B5 from \citet{Kir06}. We found the value of free-fall time using values in the Mass column and R$_{eff}$ in Table \ref{Starless}. Class 0 sources and other YSOs are obtained from \citet{Sad14} and \citet{Eva09}, respectively.

%%%%%%%%%%%
%REFERENCES%%
%%%%%%%%%%%

\begin{deluxetable}{cccccc}
\tabletypesize{\scriptsize}
\tablecolumns{6}
\tablewidth{0pc}
\tablecaption{Starless Cores in Each Clump\label{Starless}}
\tablehead{
\colhead{ID}& \colhead{Name}&\colhead{R.A. (deg)}&\colhead{Decl. (deg)}&\colhead{Mass ({\it M}$_{\odot}$)}&\colhead{R$_{eff}$ (pc)}
%\colhead{}&\colhead{}&\colhead{(deg)}&\colhead{(deg)}&\colhead{({\it M}$_{\odot}$)}&\colhead{pc}
}
\startdata
\hline
\multicolumn{6}{c}{L1448} \\
\hline
1&	J032525.4+304508&	051.3558&	+30.7522&	1.164&		2.8e-02\\
%\vspace{0.1cm}\\
\hline
\multicolumn{6}{c}{L1455} \\
\hline
1&	J032703.2+301513&	051.7633&	+30.2536&	1.421&		4.1e-02\\
2&	J032739.7+301211&	051.9154&	+30.2031&	1.660&		3.3e-02\\
3&	J032746.6+301204&	051.9442&	+30.2011	&1.558&		3.4e-02\\

\hline
\multicolumn{6}{c}{B1} \\
\hline
1&	J033158.5+304700&	052.9937&	+30.7833&	1.489&		4.3e-02\\
2&	J033217.6+304947&	053.0733&	+30.8297&	7.223&		5.5e-02\\
3&	J033214.9+305435&	053.0621&	+30.9097&	1.541&		4.4e-02\\
4&	J033243.5+305953&	053.1812&	+30.9981&	4.810&		6.9e-02\\
5&	J033243.6+310002&	053.1817&	+31.0006&	3.441&		5.9e-02\\
6&	J033249.7+310114&	053.2071&	+31.0206&	0.394&		2.4e-02\\
7&	J033251.1+310156&	053.2129&	+31.0322&	1.301&		4.1e-02\\
8&	J033301.8+310420&	053.2575&	+31.0722&	6.590&		7.5e-02\\
9&	J033305.6+310502&	053.2733&	+31.0839&	5.426&		6.5e-02\\
10&	J033305.1+310638&	053.2712&	+31.1106&	1.592&		4.3e-02\\
11&	J033311.7+310956&	053.2987&	+31.1656&	0.514&		2.6e-02\\
12&	J033318.2+310608&	053.3258&	+31.1022&	1.352&		3.9e-02\\
13&	J033325.7+310543&	053.3571&	+31.0953&	1.147&		3.6e-02\\
14&	J033301.0+312044&	053.2542&	+31.3456&	2.140&		5.0e-02\\
15&	J033327.3+311955&	053.3637&	+31.3319&	0.582&		2.8e-02\\
\hline
\multicolumn{6}{c}{IC 348} \\
\hline
1&	J034146.4+315720&	055.4433 &	+31.9556&  	0.753 & 		3.1e-02\\ 
2&	J034156.8+315844&	055.4867 &	+31.9789&   0.205 & 		1.7e-02 \\
3&	J034234.5+315620&	055.6437 &	+31.9389&   	0.462&  		2.5e-02\\
4&	J034246.3+315837&	055.6929& 	+31.9769&   	4.622& 		7.9e-02\\
5&	J034338.3+320308&	055.9096 &	+32.0522&   	2.020&  		4.2e-02\\
6&	J034342.5+320320&	055.9271 &	+32.0556&   	2.277&  		3.9e-02\\
7&	J034344.0+320250&	055.9333 &	+32.0472&   	2.876&  		4.2e-02\\
8&	J034345.8+320138&	055.9408 &	+32.0272&   	0.428&  		2.2e-02\\
9&	J034347.7+320214&	055.9487 &	+32.0372&   	0.770&  		2.9e-02\\
10&	J034405.2+320045&	056.0217 &	+32.0125&   	0.308&  		2.1e-02\\
11&	J034414.4+315753&	056.0600 &	+31.9647&   	1.660&  		4.4e-02\\
12&	J034429.7+320032&	056.1237 &	+32.0089&   	0.736&  		3.2e-02\\
13&	J034435.9+320056&	056.1496 &	+32.0156&   	1.147&  		3.7e-02\\
14&	J034436.5+315848&	056.1521 &	+31.9800&   	2.482&  		4.9e-02\\
15&	J034454.2+320020&	056.2258 &	+32.0056&   	0.907&  		3.4e-02\\
16&	J034459.9+320032&	056.2496 &	+32.0089&   	1.267&  		3.9e-02\\
17&	J034507.0+320032&	056.2792 &	+32.0089&   	0.839&  		3.3e-02\\
18&	J034358.1+320221&	055.9921 &	+32.0392&   	0.907&  		2.6e-02 \\
19&	J034358.1+320403&	055.9921 &	+32.0675&   	4.211&  		5.7e-02 \\
20&	J034448.6+320032&	056.2025 &	+32.0089&   	2.157&  		4.4e-02 \\
21&	J034516.5+320449&	056.3187 &	+32.0803&   	1.113&  		3.6e-02 \\
22&	J034520.3+320507&	056.3346 &	+32.0853&   	1.284&  		4.2e-02 \\
23&	J034050.4+314832&	055.2100 &	+31.8089  & 	0.856  &		3.3e-02 \\ 
24&	J034217.5+314620&	055.5729 	&+31.7722   &	0.325  	&	2.2e-02 \\ 
25&	J034223.1+314550&	055.5962 	&+31.7639   &	0.308  	&	2.1e-02\\ 
26&	J034207.1+314720&	055.5296 	&+31.7889   &	2.088  	&	5.2e-02 \\ 
\hline
\multicolumn{6}{c}{NGC 1333} \\
\hline
1&	J032925.4+312818&	052.3558&	+31.4717&	0.941& 	  	2.8e-02 \\
2&	J032918.3+312512&	052.3262&	+31.4200&	7.497 &	  	7.2e-02 \\
3&	J032907.4+312155&	052.2808&	+31.3653&	5.905 &	  	5.1e-02 \\
4&	J032914.9+312030&	052.3121&	+31.3417&	2.722 &	  	4.9e-02 \\
5&	J032859.5+312131&	052.2479&	+31.3586&	7.446 &	  	5.3e-02 \\
6&	J032901.3+312031&	052.2554&	+31.3419&	15.080&	  	5.9e-02 \\
7&	J032835.9+310456&	052.1496&	+31.0822&	1.181 &	  	3.6e-02\\
8&	J032832.6+310456&	052.1358&	+31.0822&	1.763 &	  	4.2e-02 \\
9&	J032842.4+310614&	052.1767&	+31.1039&	2.259 &	  	4.3e-02 \\
10&	J032847.1+310907&	052.1962&	+31.1519&	0.822 &	  	3.2e-02 \\
11&	J032829.4+310956&	052.1225&	+31.1656&	1.643 &	  	4.5e-02 \\
12&	J032831.3+311420&	052.1304&	+31.2389&	0.188 &	  	1.6e-02\\
13&	J032836.9+311326&	052.1537&	+31.2239&	3.663 &	  	4.8e-02 \\
14&	J032926.4+310730&	052.3600&	+31.1250&	0.394 &	  	2.2e-02 \\
15&	J032923.1+311000&	052.3462&	+31.1667&	0.650 &	  	2.9e-02 \\
16&	J032919.0+311136&	052.3292&	+31.1933&	3.458 &	  	5.4e-02 \\
17&	J032910.1+311331&	052.2921&	+31.2253&	24.529&	  	4.5e-02 \\
18&	J032855.2+311437&	052.2300&	+31.2436&	12.393&	  	6.2e-02 \\
19&	J032908.7+311513&	052.2862&	+31.2536&	8.096 &	  	5.5e-02 \\
20&	J032906.4+311537&	052.2767&	+31.2603&	6.505 &	  	4.1e-02 \\
21&	J032908.3+311707&	052.2846&	+31.2853&	2.773 &	  	4.2e-02 \\
22&	J032906.9+311725&	052.2787&	+31.2903&	1.541 &	  	2.7e-02 \\
23&	J032852.9+311825&	052.2204&	+31.3069&	0.531 &	  	2.6e-02 \\
24&	J032855.7+311919&	052.2321&	+31.3219&	0.993 &	  	3.2e-02 \\
\enddata
\end{deluxetable}

\begin{deluxetable}{cccccc}
\tabletypesize{\scriptsize}
\tablecolumns{6}
\tablewidth{0pc}
\tablecaption{Protostellar Cores in Each Clump\label{Protostellar}}
\tablehead{
\colhead{ID}& \colhead{Name}&\colhead{R.A.}&\colhead{Decl.}&\colhead{Mass}&\colhead{R$_{eff}$}\\
\colhead{}&\colhead{}&\colhead{(deg)}&\colhead{(deg)}&\colhead{({\it M}$_{\odot}$)}&\colhead{pc}
}
\startdata
\hline
\multicolumn{6}{c}{B5} \\
\hline
1	&	03:47:45.3&	+32:52:43.4&	859.6&	776.0\\
\hline
\multicolumn{6}{c}{L1448} \\
\hline
1&	J032536.1+304514& 051.4004& +30.7539&   17.271 & 5.6e-02\\
2&	J032538.9+304402 &	051.4121& +30.7339&   4.930 & 3.7e-02\\
3&	J032522.2+304514& 051.3425& +30.7539&   3.646 & 3.9e-02\\
\hline
\multicolumn{6}{c}{L1455} \\
\hline
1&	J032738.3+301353&	051.9096& +30.2314& 0.496  &	2.1e-02\\
2&	J032739.2+301259&	051.9133& +30.2164& 1.968  &	3.0e-02\\
3&	J032742.9+301228&	051.9287& +30.2078& 2.105  &	3.5e-02\\
4&	J032748.0+301210&	051.9500& +30.2028& 0.531  &	1.8e-02\\
\hline
\multicolumn{6}{c}{B1} \\
\hline
1&	J033228.7+310227& 053.1196& +31.0408& 0.822&	 	3.3e-02\\
2&	J033313.2+311956 &	053.3050 &	+31.3322& 2.140&	 	4.3e-02\\
3&	J033315.9+310656 &	053.3162 &	+31.1156& 14.618&	 	6.9e-02\\
4&	J033316.4+310750 &	053.3183 &	+31.1306& 6.162&	 	5.4e-02\\
5&	J033317.8+310932 &	053.3242 &	+31.1589& 17.802&	 	8.9e-02\\
6&	J033321.0+310732 &	053.3375 &	+31.1256& 17.511&	 	6.5e-02\\
7&	J033327.1+310707 &	053.3629 &	+31.1186& 3.030&	 	5.0e-02\\
8&	J033120.7+304531 &  052.8362 &	+30.7586& 3.355 & 4.4e-02\\
\hline
\multicolumn{6}{c}{IC 348} \\
\hline
1&	J034356.7+320051&  	055.9862& +32.0142& 10.014 & 5.9e-02\\ 
2&	J034357.2+320303 & 055.9883 &	+32.0508& 6.898  &	4.7e-02 \\
3&	J034401.4+320157  &	056.0058 &	+32.0325& 3.406  &	4.2e-02 \\
4&	J034402.8+320227  &	056.0117 &	+32.0408& 3.252  &	3.9e-02 \\
5&	J034406.1+320215  &	056.0254 &	+32.0375& 2.653  &	4.1e-02 \\
6&	J034412.7+320133  &	056.0529 &	+32.0258& 0.051 & 8.2e-03 \\
7&	J034421.0+315923  &	056.0875 &	+31.9897& 2.277  &	5.3e-02 \\
8&	J034443.9+320132  &	056.1829 &	+32.0256& 3.509  &	4.9e-02 \\
9&	J034351.0+320321 &	055.9625 &	+32.0558& 6.077  &	5.8e-02 \\
\hline
\multicolumn{6}{c}{NGC 1333} \\
\hline
1&	J032832.2+311108&	052.1342 &	+31.1856& 2.003&  	4.7e-02 \\
2&	J032832.6+310044&	052.1358 &	+31.0122& 0.496 & 2.6e-02 \\
3&	J032834.5+310702&	052.1437 &	+31.1172& 0.359  &	1.8e-02 \\
4&	J032839.2+310556&	052.1633 &	+31.0989& 3.132  &	5.0e-02 \\
5&	J032839.3+311826&	052.1637 &	+31.3072& 5.820  &	6.3e-02 \\
6&	J032839.8+311750&	052.1658 &	+31.2972& 3.560  &	4.3e-02 \\
7&	J032845.2+310549&	052.1883 &	+31.0969& 1.352  &	3.8e-02 \\
8&	J032900.3+311201&	052.2512 &	+31.2003& 0.753  &	2.6e-02 \\
9&	J032903.6+311455&	052.2650 &	+31.2486& 7.497  &	5.1e-02 \\
10&	J032904.6+311843&	052.2692 &	+31.3119& 0.633  &	2.4e-02 \\
11&	J032910.2+312143&	052.2925 &	+31.3619& 5.580  &	5.1e-02 \\
12&	J032910.7+311824&	052.2946 &	+31.3067& 10.818  &	6.1e-02 \\
13&	J032912.0+311306&	052.3000 &	+31.2183& 15.200  &	4.9e-02 \\
14&	J032913.4+311354&	052.3058 &	+31.2317& 4.605  &	4.1e-02 \\
15&	J032917.4+312748&	052.3225 &	+31.4633& 3.286  &	5.5e-02 \\
16&	J032918.7+312312&	052.3279 &	+31.3867& 1.386  &	2.9e-02 \\
17&	J032919.7+312348&	052.3321 &	+31.3967& 3.098  &	4.6e-02 \\
18&	J032951.4+313904&	052.4642 &	+31.6511& 1.404  &	3.3e-02 \\
\enddata
\end{deluxetable}

\begin{deluxetable}{cccc}
\tabletypesize{\scriptsize}
\tablecolumns{4}
\tablewidth{0pc}
\tablecaption{Class 0 in Each Clump}
\tablehead{
\colhead{ID}& \colhead{Source}&\colhead{R.A.}&\colhead{Decl.}\\
\colhead{}&\colhead{}&\colhead{(J2000)}&\colhead{(J2000)}}
\startdata
\hline
\multicolumn{4}{c}{L1448} \\
\hline
1&	West9 &	3:25:22.3& 30:45:10 \\
2&	West25 &	3:25:35.4& 30:45:32 \\
3&	West8 &	3:25:36.2& 30:45:17 \\
4&	West4 &	3:25:38.7& 30:44:02 \\
\hline
\multicolumn{4}{c}{L1455} \\
\hline
1&	West18 &	3:27:43.1& 30:12:26 \\
\hline
\multicolumn{4}{c}{B1} \\
\hline
1&	West17 &	3:31:20.6& 30:45:29\\
2&	West26 &	3:32:17.7 &	30:49:46\\
3&	West50 &	3:33:14.3 &	31:07:09\\
4&	West34 &	3:33:16.2 &	31:06:51\\
5&	West12 &	3:33:17.7 &	31:09:30\\
6&	West41 &	3:33:21.3 &	31:07:27\\
\hline
\multicolumn{4}{c}{IC 348} \\
\hline
1&	East11 &	3:43:50.6 &	32:03:24 \\
2&	East9 &	3:43:50.6 &	32:03:08 \\
3&	East4 &	3:43:56.4 &	32:00:49 \\
4&	East5 &	3:43:56.6 &	32:03:03 \\
5&	East17 &	3:44:02.1 &	32:02:02\\
\hline
\multicolumn{4}{c}{NGC 1333} \\
\hline
1&	West162 &	3:28:38.6 &	31:06:00 \\
2&	West33 &	3:29:00.4 &	31:11:57 \\
3&	West19 &	3:29:01.8 &	31:15:34 \\
4&	West40 &	3:29:03.9 &	31:14:43 \\
5&	West87 &	3:29:06.7 &	31:15:33 \\
6&	West6 &	3:29:10.3 &	31:13:28 \\
7&	West14 &	3:29:11.0 &	31:18:26 \\
8&	West13 &	3:29:11.9 &	31:13:05 \\
9&	West30 &	3:29:13.5 &	31:13:54 \\
10&	West23 &	3:29:17.2 &	31:27:43 \\
11&	West37 &	3:29:18.8 &	31:23:12 \\
12&	West28 &	3:29:51.7 &	31:39:03 \\
\enddata
\end{deluxetable}

\begin{deluxetable}{ccccc}
\tabletypesize{\scriptsize}
\tablecolumns{5}
\tablewidth{0pc}
\tablecaption{Class I in Each Clump\label{ClassI}}
\tablehead{
\colhead{ID}& \colhead{{\it Spitzer} Source Name}&\colhead{c2d}&\colhead{R.A.}&\colhead{Decl.}\\
\colhead{}&\colhead{(JHHMMSS.ss+DDMMSS.s)}&\colhead{classification}&\colhead{(deg)}&\colhead{(deg)}
}
\startdata
\hline
\multicolumn{5}{c}{B5} \\
\hline
1&	J034705.43+324308.5	 &     YSOc\_red  &           56.77262&	32.71903\\
2&	J034741.58+325144.1	  &      YSOc      &          56.92325&	32.86225\\
\hline
\multicolumn{5}{c}{L1448} \\
\hline
1&	J032539.12+304358.2&	      YSOc\_red &              51.413&	30.73283\\
2&	J032538.83+304406.2	&       YSOc\_red     &          51.41179&	30.73506\\
3&	J032536.22+304515.7	&      red           &         51.40092&	30.75436\\
4&	J032536.49+304522.2	 &     YSOc\_red       &        51.40204&	30.75617\\
5&	J032522.32+304513.9	  &    YSOc\_red        &       51.343	&	30.75386\\
\hline
\multicolumn{5}{c}{L1455} \\
\hline
1&	J032800.39+300801.3&	      YSOc\_red      &         52.00162&	30.13369\\
2&	J032738.83+301257.9	&       YSOc\_red          &     51.91179&	30.21608\\
3&	J032739.08+301303.1	&      YSOc\_red           &    51.91283&	30.21753\\

\hline
\multicolumn{5}{c}{B1} \\
\hline
1&	J033120.98+304530.1	 &     YSOc\_red            &   52.83742&	30.75836\\
2&	J033257.84+310608.3	   &   YSOc\_red              & 53.241&	31.10231\\
3&	J033309.56+310531.2	    &  YSOc\_PAH-em            &53.28983&	31.092\\
4&	J033313.80+312005.3	     & YSOc\_red     &          53.3075&	31.33481\\
5&	J033316.44+310652.5&	      red         &           53.3185&	31.11458\\
6&	J033316.65+310755.2	&       YSOc\_red        &       53.31937&	31.132\\
7&	J033320.32+310721.5	 &     YSOc\_red          &     53.33467&	31.12264\\
8&	J033327.29+310710.2	  &    YSOc\_red           &    53.36371&	31.1195\\
\hline
\multicolumn{5}{c}{IC 348} \\
\hline
1&	J034329.43+315219.5&	      YSOc\_star+dust(IR1)&    55.87262&	31.87208\\
2&	J034356.52+320052.8	&      red                    &55.9855	&32.01467\\
3&	J034356.84+320304.7	 &     YSOc\_red               &55.98683	&	32.05131\\
4&	J034402.40+320204.9	  &    YSOc\_red               &56.01	&32.03469\\
5&	J034409.20+320237.8	   &   YSOc\_star+dust(IR1)    &56.03833	&	32.04383\\
6&	J034421.35+315932.6	    &  YSOc\_red               &56.08896	&	31.99239\\
7&	J034443.32+320131.5	    &  YSOc\_red               &56.1805	&32.02542\\
8&	J034158.67+314821.4&	      YSOc\_red   &            55.49446&	31.80594\\
9&	J034202.17+314802.1	&       YSOc\_red      &         55.50904	&	31.80058\\
\hline
\multicolumn{5}{c}{NGC 1333} \\
\hline
1&	J032832.56+311105.1&	      YSOc\_red         &      52.13567&	31.18475\\
2&	J032834.49+310051.1	&       YSOc\_star+dust(IR1)  &  52.14371	&	31.01419\\
3&	J032834.53+310705.5	&      YSOc                  & 52.14387	&	31.11819\\
4&	J032837.09+311330.8	 &     YSOc\_red          &     52.15454	&	31.22522\\
5&	J032839.71+311731.9	  &    YSOc\_red            &   52.16546	&	31.29219\\
6&	J032840.63+311756.5	   &   red                &    52.16929	&	31.29903\\
7&	J032843.28+311732.9	    &  YSOc\_star+dust(IR1)  &  52.18033	&	31.29247\\
8&	J032845.30+310541.9	     & YSOc\_red              & 52.18875	&	31.09497\\
9&	J032851.26+311739.3&	      YSOc\_star+dust(IR2) &   52.21358	&	31.29425\\
10&	J032855.55+311436.7	&       rising                & 52.23146	&	31.24353\\
11&	J032857.36+311415.9	&      YSOc\_red              & 52.239	&31.23775\\
12&	J032858.43+312217.5	 &     YSOc\_red              & 52.24346	&	31.37153\\
13&	J032900.55+311200.8	  &    red                   & 52.25229	&	31.20022\\
14&	J032901.56+312020.6	   &   rising                & 52.2565	&31.33906\\
15&	J032903.33+312314.6	    &  YSOc\_red              & 52.26387	&	31.38739\\
16&	J032903.78+311603.8	     & rising                & 52.26575	&	31.26772\\
17&	J032904.06+311446.5&	      YSOc\_red            &   52.26692	&	31.24625\\
18&	J032907.78+312157.3	&       rising                & 52.28242	&	31.36592\\
19&	J032909.10+312128.7	&      YSOc      &             52.28792	&31.35797\\
20&	J032910.65+311340.0	 &     YSOc\_red&               52.29437	&	31.22778\\
21&	J032910.68+311820.6	  &    YSOc\_red &              52.2945	&31.30572\\
22&	J032910.99+311826.0	   &   YSOc\_red  &             52.29579	&	31.30722\\
23&	J032911.26+311831.4	    &  YSOc\_red   &            52.29692	&	31.30872\\
24&	J032912.06+311305.4	     & YSOc\_red    &           52.30025	&	31.21817\\
25&	J032912.97+311814.3&	      YSOc      &             52.30404	&	31.30397\\
26&	J032913.54+311358.2	&       red           &         52.30642	&	31.23283\\
27&	J032917.17+312746.5	&      YSOc\_red       &        52.32154	&	31.46292\\
28&	J032918.26+312319.9	 &     YSOc\_red        &       52.32608	&	31.38886\\
29&	J032923.48+313329.5	  &    YSOc\_red         &      52.34783	&	31.55819\\
30&	J032951.82+313906.0	   &   red               &     52.46592	&	31.65167\\
\enddata
\end{deluxetable}

\begin{deluxetable}{ccccc}
\tabletypesize{\scriptsize}
\tablecolumns{5}
\tablewidth{0pc}
\tablecaption{Class Flat in Each Clump\label{ClassFlat}}
\tablehead{
\colhead{ID}& \colhead{{\it Spitzer} Source Name}&\colhead{c2d}&\colhead{R.A.}&\colhead{Decl.}\\
\colhead{}&\colhead{(JHHMMSS.ss+DDMMSS.s)}&\colhead{classification}&\colhead{(deg)}&\colhead{(deg)}
}
\startdata
\hline
\multicolumn{5}{c}{L1455} \\
\hline
1&	J032747.67+301204.5	& YSOc\_star+dust(IR1)&	51.94862&	30.20125\\
2&	J032835.03+302009.9	&	YSOc\_star+dust(IR4) &	52.14596&	30.33608\\

\hline
\multicolumn{5}{c}{B1} \\
\hline
1&	J033229.17+310240.8& YSOc\_star+dust(IR1)&	53.12154&	31.04467\\
2&	J033306.41+310804.6& YSOc\_star+dust(IR1)&	53.27671&	31.13461\\
3&	J033312.84+312124.2& YSOc\_star+dust(IR1)&	53.3035	&	31.35672\\

\hline
\multicolumn{5}{c}{IC 348} \\
\hline
1&	J034256.05+315644.8&	YSOc               & 55.73354&	31.94578\\
2&	J034336.02+315009.0&	YSOc\_star+dust(IR1) &	55.90008&	31.83583\\
3&	J034345.17+320358.6&	YSOc\_star+dust(IR1) &	55.93821&	32.06628\\
4&	J034359.41+320035.7&	YSOc\_red           & 55.99754&	32.00992\\
5&	J034412.98+320135.5&	YSOc               & 56.05408&	32.02653\\
6&	J034435.34+322837.2&	YSOc\_red           & 56.14725&	32.477	\\
7&	J034141.09+314804.6&	YSOc       &         55.42121&	31.80128\\

\hline
\multicolumn{5}{c}{NGC 1333} \\
\hline
1&	J032838.78+311806.6& YSOc\_star+dust(IR2)&	52.16158&	31.30183\\
2&	J032848.77+311608.8&	YSOc\_red           & 52.20321&	31.26911\\
3&	J032856.60+310737.0&	YSOc\_red           & 52.23583&	31.12694\\
4&	J032858.26+312209.2&	YSOc\_star+dust(IR2) &	52.24275&	31.36922\\
5&	J032859.23+312032.5&	YSOc\_star+dust(IR1) &	52.24679&	31.34236\\
6&	J032859.32+311548.7&	YSOc\_star+dust(IR4) &	52.24717&	31.26353\\
7&	J032901.88+311653.2&	YSOc\_star+dust(IR2) &	52.25783&	31.28144\\
8&	J032904.95+312038.4&	YSOc\_star+dust(IR2) &	52.27062&	31.344	\\
9&	J032911.89+312127.0&	YSOc               & 52.29954&	31.3575	\\
10&	J032912.06+311301.7&	YSOc\_red           & 52.30025&	31.21714\\
11&	J032918.67+312017.7& YSOc\_star+dust(IR2)	&	52.32779&	31.33825\\
12&	J032920.06+312407.5& YSOc\_star+dust(IR2)	&	52.33358&	31.40208\\
13&	J032920.44+311834.2& YSOc\_star+dust(IR1)	&	52.33517&	31.3095	\\
14&	J032924.09+311957.6& YSOc\_star+dust(IR2)	&	52.35037&	31.33267\\
\enddata
\end{deluxetable}

\begin{deluxetable}{ccccc}
\tabletypesize{\scriptsize}
\tablecolumns{5}
\tablewidth{0pc}
\tablecaption{Class II in Each Clump\label{ClassII}}
\tablehead{
\colhead{ID}& \colhead{{\it Spitzer} Source Name}&\colhead{c2d}&\colhead{R.A.}&\colhead{Decl.}\\
\colhead{}&\colhead{(JHHMMSS.ss+DDMMSS.s)}&\colhead{classification}&\colhead{(deg)}&\colhead{(deg)}
}
\startdata
\hline
\multicolumn{5}{c}{L1455} \\
\hline
1&	J032738.25+301358.6	& YSOc\_star+dust(IR4)&	51.90937&	30.23294\\
2&	J032800.09+300847.0	& YSOc\_star+dust(IR2)&	52.00037&	30.14639\\

\hline
\multicolumn{5}{c}{B1} \\
\hline
1&	J033232.99+310221.7& YSOc\_star+dust(IR1)&	53.13746&	31.03936\\
2&	J033234.05+310055.8& YSOc\_star+dust(IR2)&	53.14187&	31.0155	\\
3&	J033247.20+305916.3& YSOc\_star+dust(IR2)&	53.19667&	30.98786\\
4&	J033330.41+311050.6& YSOc\_star+dust(IR1)&	53.37671&	31.18072\\
5&	J033341.29+311341.0& YSOc\_star+dust(IR1)&	53.42204&	31.22806\\

\hline
\multicolumn{5}{c}{IC 348} \\
\hline
1&	J034244.50+315958.7& YSOc\_star+dust(IR1)&	55.68542&	31.99964\\
2&	J034249.18+315011.2	&YSOc\_star+dust(IR4) &	55.70492&	31.83644\\
3&	J034255.95+315842.0	&YSOc\_star+dust(IR1) &	55.73312&	31.97833\\
4&	J034313.70+320045.2	&YSOc\_star+dust(IR2) &	55.80708&	32.01256\\
5&	J034325.48+315516.5 &YSOc\_star+dust(MP1)	&	55.85617&	31.92125\\
6&	J034328.21+320159.1	&YSOc\_star+dust(IR2) &	55.86754&	32.03308\\
7&	J034355.00+320103.1	&YSOc\_star+dust(IR2) &	55.97917&	32.01753\\
8&	J034355.24+315532.1	&YSOc\_star+dust(IR1) &	55.98017&	31.92558\\
9&	J034356.03+320213.3	&YSOc\_star+dust(IR2) &	55.98346&	32.03703\\
10&	J034357.23+320133.7& YSOc\_star+dust(IR2)	&	55.98846&	32.02603\\
11&	J034359.65+320154.1	&YSOc               & 55.99854&	32.03169\\
12&	J034400.48+320432.7	&YSOc\_red           & 56.002	&	32.07575\\
13&	J034405.78+320001.1 &YSOc\_star+dust(IR1)	&	56.02408&	32.00031\\
14&	J034405.78+320028.5	&YSOc               & 56.02408&	32.00792\\
15&	J034410.13+320404.5 &YSOc\_star+dust(IR2)	&	56.04221&	32.06792\\
16&	J034411.63+320313.1 &YSOc\_star+dust(IR3)	&	56.04846&	32.05364\\
17&	J034415.84+315936.7 &YSOc\_star+dust(IR2)	&	56.066	&	31.99353\\
18&	J034418.17+320457.0 &YSOc\_star+dust(IR3)	&	56.07571&	32.0825\\
19&	J034419.25+320734.7 &YSOc\_star+dust(IR4)	&	56.08021&	32.12631\\
20&	J034420.18+320856.5 &YSOc\_star+dust(IR2)	&	56.08408&	32.14903\\
21&	J034421.23+320114.5 &YSOc\_star+dust(IR4)	&	56.08846&	32.02069\\
22&	J034422.29+320542.8 &YSOc\_star+dust(IR3)	&	56.09287&	32.09522\\
23&	J034422.58+320153.6 &YSOc\_star+dust(IR4)	&	56.09408&	32.03156\\
24&	J034424.46+320143.7 &YSOc\_star+dust(IR2)	&	56.10192&	32.02881\\
25&	J034425.55+320617.1 &YSOc\_star+dust(IR4)	&	56.10646&	32.10475\\
26&	J034426.04+320430.4 &YSOc\_star+dust(IR1)	&	56.1085	&	32.07511\\
27&	J034426.70+320820.3 &YSOc\_star+dust(IR2)	&	56.11125&	32.13897\\
28&	J034428.51+315954.0 &YSOc\_star+dust(IR3)	&	56.11879&	31.99833\\
29&	J034428.95+320137.9 &YSOc\_star+dust(IR2)	&	56.12062&	32.02719\\
30&	J034429.80+320054.6 &YSOc\_star+dust(IR2)	&	56.12417&	32.01517\\
31&	J034431.19+320558.9 &YSOc\_star+dust(IR4)	&	56.12996&	32.09969\\
32&	J034431.37+320014.2 &YSOc\_star+dust(IR2)	&	56.13071&	32.00394\\
33&	J034433.79+315830.2 &YSOc\_star+dust(IR4)	&	56.14079&	31.97506\\
34&	J034434.81+315655.2 &YSOc\_star+dust(IR4)	&	56.14504&	31.94867\\
35&	J034445.20+320119.6 &YSOc\_star+dust(IR1)	&	56.18833&	32.02211\\
36&	J034124.42+315327.9& YSOc\_star+dust(IR1)&	55.35175&	31.89108\\
37&	J034153.26+315019.2& YSOc\_star+dust(IR1)&	55.47192&	31.83867\\
38&	J034201.01+314913.4& YSOc\_star+dust(IR3)&	55.50421&	31.82039\\
39&	J034204.34+314711.6& YSOc\_star+dust(IR1)&	55.51808&	31.78656\\
40&	J034210.69+314705.6& YSOc\_star+dust(IR3)&	55.54454&	31.78489\\
41&	J034301.94+314435.6& YSOc\_star+dust(IR1)&	55.75808&	31.74322\\
42&	J034321.47+314246.3& YSOc\_star+dust(IR3)&	55.83946&	31.71286\\

\hline
\multicolumn{5}{c}{NGC 1333} \\
\hline
1&	J032844.09+312052.7& YSOc\_star+dust(IR2)&	52.18371&	31.34797\\
2&	J032847.65+312406.0 &YSOc\_star+dust(IR1)&	52.19854&	31.40167\\
3&	J032847.84+311655.1 &YSOc\_star+dust(IR2)&	52.19933&	31.28197\\
4&	J032851.03+311818.5 &YSOc\_star+dust(IR2)&	52.21262&	31.30514\\
5&	J032851.08+311632.4 &YSOc\_star+dust(IR2)&	52.21283&	31.27567\\
6&	J032851.20+311954.8 &YSOc\_star+dust(IR2)&	52.21333&	31.33189\\
7&	J032852.15+311547.1 &YSOc\_star+dust(IR2)&	52.21729&	31.26308\\
8&	J032852.17+312245.3 &YSOc\_star+dust(IR3)&	52.21737&	31.37925\\
9&	J032852.92+311626.4 &YSOc\_star+dust(IR2)&	52.2205	&	31.274	\\
10&	J032853.96+311809.3 &YSOc\_star+dust(IR3)&	52.22483&	31.30258\\
11&	J032854.09+311654.2 &YSOc\_star+dust(IR3)&	52.22537&	31.28172\\
12&	J032854.63+311651.1 &YSOc\_star+dust(IR2)&	52.22762&	31.28086\\
13&	J032855.08+311628.7 &YSOc\_star+dust(IR1)&	52.2295	&	31.27464\\
14&	J032856.12+311908.4	&YSOc               & 52.23383&	31.319	\\
15&	J032856.32+312227.9 &YSOc\_star+dust(IR2)&	52.23467&	31.37442\\
16&	J032856.65+311835.5 &YSOc\_star+dust(IR2)&	52.23604&	31.30986\\
17&	J032856.97+311622.3 &YSOc\_star+dust(IR4)&	52.23737&	31.27286\\
18&	J032857.18+311534.6 &YSOc\_star+dust(IR2)&	52.23825&	31.25961\\
19&	J032858.27+312202.0 &YSOc\_star+dust(IR1)&	52.24279&	31.36722\\
20&	J032859.56+312146.7 &YSOc\_star+dust(IR1)&	52.24817&	31.36297\\
21&	J032902.81+312217.2 &YSOc\_star+dust(IR2)&	52.26171&	31.37144\\
22&	J032903.15+312238.0 &YSOc\_star+dust(IR1)&	52.26312&	31.37722\\
23&	J032903.22+312545.1 &YSOc\_star+dust(IR2)&	52.26342&	31.42919\\
24&	J032903.87+312148.6 &YSOc\_star+dust(IR1)&	52.26612&	31.3635	\\
25&	J032904.68+311659.0	&YSOc               & 52.2695	&	31.28306\\
26&	J032904.73+311134.9	&YSOc               & 52.26971&	31.19303\\
27&	J032905.18+312036.9	&YSOc\_red           & 52.27158&	31.34358\\
28&	J032906.33+311346.4 &YSOc\_star+dust(IR3)&	52.27637&	31.22956\\
29&	J032907.96+312251.4 &YSOc\_star+dust(IR2)&	52.28317&	31.38094\\
30&	J032909.34+312104.1 &YSOc\_star+dust(IR3)&	52.28892&	31.35114\\
31&	J032909.40+311413.8	&YSOc               & 52.28917&	31.23717\\
32&	J032909.49+312720.9 &YSOc\_star+dust(IR2)&	52.28954&	31.45581\\
33&	J032909.65+312256.3 &YSOc\_star+dust(IR4)&	52.29021&	31.38231\\
34&	J032910.47+312334.7 &YSOc\_star+dust(IR2)&	52.29362&	31.39297\\
35&	J032910.84+311642.6 &YSOc\_star+dust(IR1)&	52.29517&	31.2785	\\
36&	J032913.14+312252.8 &YSOc\_star+dust(IR2)&	52.30475&	31.38133\\
37&	J032914.40+311444.1	&YSOc               & 52.31	&	31.24558\\
38&	J032916.61+312349.4 &YSOc\_star+dust(IR2)&	52.31921&	31.39706\\
39&	J032916.83+312325.1 &YSOc\_star+dust(IR2)&	52.32012&	31.39031\\
40&	J032917.68+312245.0 &YSOc\_star+dust(IR1)&	52.32367&	31.37917\\
41&	J032917.78+311948.0 &YSOc\_star+dust(IR2)&	52.32408&	31.33	\\
42&	J032918.74+312325.4 &YSOc\_star+dust(IR2)&	52.32808&	31.39039\\
43&	J032921.57+312110.3 &YSOc\_star+dust(IR2)&	52.33987&	31.35286\\
44&	J032921.87+311536.2 &YSOc\_star+dust(IR2)&	52.34112&	31.26006\\
45&	J032923.17+312030.2 &YSOc\_star+dust(IR2)&	52.34654&	31.34172\\
46&	J032923.25+312653.1 &YSOc\_star+dust(IR2)&	52.34687&	31.44808\\
47&	J032925.93+312640.1	&YSOc              	& 52.35804&	31.44447\\
48&	J032929.27+311834.7 &YSOc\_star+dust(IR4)&	52.37196&	31.30964\\
49&	J032929.80+312102.6 &YSOc\_star+dust(IR3)&	52.37417&	31.35072\\
50&	J032930.40+311903.3 &YSOc\_star+dust(IR2)&	52.37667&	31.31758\\
51&	J032932.57+312436.9 &YSOc\_star+dust(IR1)&	52.38571&	31.41025\\
52&	J032932.88+312712.6 &YSOc\_star+dust(IR2)&	52.387	&	31.4535\\
53&	J032937.73+312202.5 &YSOc\_star+dust(IR2)&	52.40721&	31.36736\\
\enddata
\end{deluxetable}

\begin{deluxetable}{ccccc}
\tabletypesize{\scriptsize}
\tablecolumns{5}
\tablewidth{0pc}
\tablecaption{Class III in Each Clump\label{ClassIII}}
\tablehead{
\colhead{ID}& \colhead{{\it Spitzer} Source Name}&\colhead{c2d}&\colhead{R.A.}&\colhead{Decl.}\\
\colhead{}&\colhead{(JHHMMSS.ss+DDMMSS.s)}&\colhead{classification}&\colhead{(deg)}&\colhead{(deg)}
}
\startdata
\hline
\multicolumn{5}{c}{IC 348} \\
\hline
1&	J034409.16+320709.3& YSOc\_star+dust(MP1)	  &   56.03817	& 32.11925\\
2&	J034430.14+320118.2	&	YSOc\_star+dust(IR4)    &  56.12558	&	32.02172\\
3&	J034223.33+315742.7	&	YSOc\_star+dust(IR4)     & 55.59721	&	31.96186\\
4&	J034436.96+320645.2	&	YSOc\_star+dust(IR4)     & 56.154	&	32.11256\\
\hline
\multicolumn{5}{c}{NGC 1333} \\
\hline
1&	J032843.24+311042.7	& YSOc\_star+dust(IR4)&	     52.18017&	31.17853\\
2&	J032857.21+311419.1	& YSOc\_star+dust(MP1)	&      52.23837	& 31.23864\\
3&	J032858.11+311803.7	& YSOc\_star+dust(MP1)	&     52.24212	& 31.30103\\
4&	J032916.69+311618.2	& YSOc\_star+dust(MP1)	 &    52.31954	& 31.27172\\
5&	J032926.81+312647.6	& YSOc\_star+dust(MP1)	  &   52.36171	& 31.44656\\
\enddata
\end{deluxetable}

%%%%%%%%%%%   BIBLIOGRAPHY  %%%%%%%%%%%%%%%%
%\bibliographystyle{apj}
%\bibliography{references}

\end{document}